# Wide therapeutic time window for nimesulide neuroprotection in a

# model of transient focal cerebral ischemia in the rat

Authors: Eduardo Candelario-Jalil [a, b, *], Armando González-Falcón [a], Michel García-Cabrera [a],

Olga Sonia León [a], Bernd L. Fiebich [b, c]

[a] *Department of Pharmacology, University of Havana (CIEB-IFAL), Apartado Postal 6079, Havana City 10600, Cuba*
[b] *Neurochemistry Research Group, Department of Psychiatry, University of Freiburg Medical School, Hauptstrasse 5, D-79104 Germany*
[c] *VivaCell Biotechnology GmbH, Ferdinand-Porsche-Str. 5, D-79211 Denzlingen, Germany*

*Author to whom all correspondence should be addressed:

Dr. Eduardo Candelario-Jalil
Neurochemistry Research Group,
Department of Psychiatry,
University of Freiburg Medical School
Hauptstrasse 5,
D-79104 Germany
Tel.: +49-761-270-6668
Fax: +49-761-270-6916
E-mail: candelariojalil@yahoo.com

**Acknowledgements:** The authors are grateful to Dr. Mayra Levi (Gautier-Bagó Laboratories) for kindly providing nimesulide for these studies. We would like to thank Noël H. Mhadu for expert technical assistance. ECJ is greatly indebted to Prof. Joe E. Springer (Department of Anatomy and Neurobiology, University of Kentucky Medical Center, Lexington, USA) for his constant support and encouragement. ECJ was supported by a research fellowship from the Alexander von Humboldt Foundation (Germany).

## Abstract

Results from several studies indicate that cyclooxygenase-2 (COX-2) is involved ischemic brain injury. The purpose of this study was to evaluate the neuroprotective effects of the selective COX-2 inhibitor nimesulide on cerebral infarction and neurological deficits in a standardized model of transient focal cerebral ischemia in rats. Three doses of nimesulide (3, 6 and 12 mg/kg; i.p.) or vehicle were administered immediately after stroke and additional doses were given at 6, 12, 24, 36 and 48 h after ischemia. In other set of experiments, the effect of nimesulide was studied in a situation in which its first administration was delayed for 3 to 24 h after ischemia. Total, cortical and subcortical infarct volumes and functional outcome (assessed by neurological deficit score and rotarod performance) were determined 3 days after ischemia. The effect of nimesulide on prostaglandin $E_2$ ($PGE_2$) levels in the injured brain was also investigated. Nimesulide dose-dependently reduced infarct volume and improved functional recovery when compared to vehicle. Of interest is the finding that neuroprotection conferred by nimesulide (reduction of infarct size and neurological deficits and improvement of rotarod performance) was also observed when treatment was delayed until 24 h after ischemia. Further, administration of nimesulide in a delayed treatment paradigm completely abolished $PGE_2$ accumulation in the postischemic brain, suggesting that COX-2 inhibition is a promising therapeutic strategy for cerebral ischemia to target the late-occurring inflammatory events which amplify initial damage.

*Theme:* Disorders of the nervous system

*Topic:* Ischemia

## 1. INTRODUCTION

Stroke is a leading cause of death and disability worldwide. Development of an effective therapeutic strategy for stroke has been a priority of neuroscientists for decades. Considerable research efforts have been devoted to the development of neuroprotective agents to save neurons from the biochemical and metabolic consequences of ischemic brain injury.

Current approaches to treat acute ischemic stroke include thrombolytic therapy (reperfusion), neuroprotection, and administration of neurorestorative agents. Intravenous thrombolysis within 3 hours after symptoms onset represents the first therapeutical approach that can effectively treat acute ischemic stroke [70]. Although adequate treatment should be started as early as possible, most patients still arrive at the hospital too late to receive the maximum benefit from this emerging therapy [2]. Despite the recently published PROACT II clinical trial, which first demonstrated the efficacy of intra-arterial thrombolysis within 6 hours of stroke onset [28], until now, thrombolysis has not been approved for intravenous administration >3 hours after stroke onset because the risk of hemorrhage increases with time [2,28].

In search of an effective treatment of stroke, numerous studies have been conducted to understand the pathophysiological mechanisms that lead to neuronal death. Among the initial events in the ischemic cascade are the widespread neuronal depolarization and massive release of glutamate (excitotoxicity) leading to calcium influx. Approaches focusing on blocking presynaptic glutamate release, ionotropic glutamate receptors, or voltage-sensitive calcium or sodium channels have so far failed to demonstrate a proven clinical efficacy [18,23,30].

Other therapeutic strategies for stroke include hypothermia [20,65,81], antioxidants [7,13,14,64], blockade of excessive synaptic $Zn^{2+}$ release [75], antiapoptotic strategies [26,53], administration of growth factors [1,40], erythropoietin [67], recombinant human interleukin-1 receptor antagonist [47], statins [73], gene therapy [80,81] and approaches involving transfer of new cells, such as stem cells, and neuronal precursors cells [1,46,63].

Recent experimental evidences have shown that brain damage occurring after focal cerebral ischemia (FCI) develops over a relatively long period of time [34,51]. One of the processes that plays a pivotal role in the delayed progression of brain damage is postischemic inflammation [19,34,51]. Cerebral ischemia is followed by infiltration of neutrophils in the injured brain, an event initiated by the expression of pro-inflammatory cytokines, chemokines, and adhesion molecules (for review, see reference 19). In addition, the marked expression of inflammation-related enzymes such as inducible nitric oxide synthase (iNOS) and cyclooxygenase-2 (COX-2) plays an important role in the secondary events that amplify cerebral damage after ischemia.

COX-2 mRNA and protein levels have shown to be significantly increased within neurons and vascular cells in both experimental animal models of cerebral ischemia [15,21,38,43,50,51,55,61] and in infarcted human brain [35,60]. Moreover, administration of highly selective COX-2 inhibitors has been proven to reduce brain damage and prostaglandin accumulation following cerebral ischemia [9-12,50,51]. The observations that enhanced COX-2 activity contributes to amplify cerebral injury after stroke offer an interesting prospect for targeting the late phase of the damage with COX-2 inhibitors.

In view that most patients with ischemic stroke reach the hospital several hours after the onset of symptoms, it is now considered of paramount importance to demonstrate significant neuroprotection after delayed administration of a compound. Hence, the aim of the present study was to investigate the effectiveness of the selective COX-2 inhibitor nimesulide (N-(4-nitro-2-phenoxyphenyl)-methanesulfonamide) when treatment started at various intervals after the onset of stroke. Nimesulide is a non-steroidal anti-inflammatory drug with potent effects, showing a high affinity and selectivity for COX-2 [16]. This compound has been widely used for 20 years to treat inflammatory conditions and fever and readily crosses the blood-brain barrier [16,69].

This study was prompted by our previous results with this COX-2 inhibitor in transient global cerebral ischemia [9,12] and in a model of in vivo kainate-induced excitotoxicity [8]. The present study used a standardized model of middle cerebral artery occlusion (MCAO) induced by an intraluminal suture and several measures of nimesulide's efficacy were considered (infarct volume, neurological deficits, motor impairment using the rotarod test, and accumulation of $PGE_2$ in infarcted cerebral cortex). We found that nimesulide potently protects against neuronal damage and improves functional outcome even when treatment is delayed up to 24 h after the onset of FCI.

## 2. MATERIALS AND METHODS

### 2.1. Animals

All the experimental procedures were performed strictly according to the regulations of the Havana University's animal ethical committee and the guidelines of the National Institutes of Health (Bethesda, MD, USA) for the care and use of laboratory animals for experimental procedures. Our institutional animal care and use committee approved the experimental protocol (No. 03/113). Male Sprague-Dawley rats (CENPALAB, Havana, Cuba) weighing 270-320 g at the time of surgery were used in the present study. The animals were quarantined for at least 7 days before the experiment. Animals were housed in groups in a room whose environment was maintained at 21-25 ºC, 45-50 % humidity and 12-h light/dark cycle. They had free access to pellet chow and water.

### 2.2. Procedures for transient middle cerebral artery occlusion (MCAO)

Rats were anesthetized with chloral hydrate (300 mg/kg body weight, i.p.). Once surgical levels of anesthesia were attained (assessed by absence of hind leg withdrawal to pinch), ischemia was induced by using an occluding intraluminal suture [41,59,80]. Briefly, the right common carotid artery (CCA) and the external carotid artery (ECA) were exposed by a ventral midline neck incision and ligated with a 3-0 silk suture. The pterygopalatine branch of the internal carotid artery was clipped to prevent incorrect insertion of the occluder filament. Arteriotomy was performed in the CCA approximately 3 mm proximal to the bifurcation and a 3-0 monofilament nylon suture (Shenzhen Runch Industrial Corp., China), whose tip had been rounded by being heated near a flame, was introduced into the internal carotid artery (ICA) until a mild resistance to this advancement indicated that the intraluminal occluder had entered the anterior cerebral artery (ACA) and occluded the origin of the ACA, the middle cerebral artery (MCA) and posterior communicating arteries. The occluding suture was kept in place for 1 h. At the end of the ischemic period, the suture was gently retracted to allow reperfusion of the ischemic region. The incision was closed and animals were allowed to recover from anesthesia and to eat and drink freely. Rectal temperature was maintained at 37±0.5 °C with a heat lamp and electrically heated mat during surgery, stroke, and reperfusion. By using this standardized procedure, we obtained large and reproducible infarcted regions involving the temporoparietal cortex and the laterocaudal part of the caudate putamen in ischemic animals. Only 1 h of ischemia was employed in the present study because findings from our pilot studies (standardization of the model) indicated that in our experimental conditions, endovascular occlusion of the middle cerebral artery for 1 h is enough to produce large infarcts in the ischemic regions. Infarct volumes obtained with our procedure are very similar to those found with longer periods of ischemia [13,51,76]. Several studies have also found large infarct areas using only 1 h of transient focal ischemia and have used this model to investigate the effects of different compounds in experimental ischemic stroke in the rat [58,67,74,80].

Since nimesulide's neuroprotection could easily result from drug-induced hypothermia rather than a specific pharmacological effect, we strictly monitored rectal temperature and found that nimesulide did not modify this physiological variable (data not shown). To allow for better postoperative recovery, we chose not to monitor physiological parameters in the present study because additional surgical procedures are needed for this monitoring. Nevertheless, we performed a separate experiment to investigate the effects of nimesulide on major physiological variables in ischemic rats. Mean arterial blood pressure, blood glucose, rectal temperature, hematocrit, blood pH and blood gases ($pO_2$ and $pCO_2$) were measured before ischemia, during the occlusion and 30 min after each vehicle or nimesulide (12 mg/kg; i.p.) administration. The effects observed with nimesulide in the present study were not related to modification of physiological variables since these parameters did not differ between nimesulide-treated and vehicle-treated rats (data not shown). These findings are in agreement with our previous results [9], suggesting that nimesulide does not significantly change major physiological variables.

### 2.3. Evaluation of nimesulide's effects on ischemic damage
### 2.3.1. Dose-response experiment
In order to evaluate the effect of different doses of nimesulide on rat FCI, nimesulide (3, 6 and 12 mg/kg) was given to rats by intraperitoneal administration starting immediately after ischemia (n=8-10 animals per group). Additional doses were given at 6, 12, 24, 36 and 48 h after stroke. This treatment schedule and dosage range were based on the pharmacokinetic profile of nimesulide [71] and on our previous experience with this compound in a model of global cerebral ischemia [9,12]. We also performed an experiment in which nimesulide was given as a single dose. Control ischemic animals received the vehicle of nimesulide (polyvinylpyrrolidone, PVP, n=11). Nimesulide was dissolved in a 2% PVP solution [8,9,12]. Neurological evaluation, motor impairment and infarct volumes were determined 3 days later as detailed below.

### 2.3.2. Effects of a single dose of nimesulide
The effect of a single dose of nimesulide (12 mg/kg; i.p.) given immediately after ischemia was examined (n=9). A single injection vehicle-treated group was also included (n=8).

### 2.3.3. Therapeutic time window assessment
After investigating the dose-response relationship, we studied the effect of the maximal effective dose of nimesulide (12 mg/kg; i.p.) in a situation in which its first administration was delayed for 3 to 24 h after ischemia (n= 9-11 animals per group; see Table 1 for detailed treatment schedules). Vehicle-treated animals, which underwent transient MCAO, were also included (n= 9-10 animals per group). The evaluated variables were measured at 3 days after ischemia because according to results from our pilot studies, brain damage is completed by 72 h in our model of transient MCAO (infarct volume at 72 h is not significantly different from that measured at 96 h post-stroke, data not shown).

### 2.4. Neurological evaluation and assessment of functional outcome
Behavioral tests were performed before MCAO (presurgery) and at 24, 48 and 72 h after MCAO. Each rat acted as its own control. Neurological deficits were assessed according to a six-point scale: 0= no neurological deficits, 1= failure to extend left forepaw fully, 2= circling to the left, 3= falling to left, 4= no spontaneous walking with a depressed level of consciousness, and 5= death [41,44].

Given that the rotarod test is one of the most sensitive to the sensorimotor defects induced by the ischemic insult [6,33,72], motor impairment in this study was assessed with the use of the accelerating rotarod (Ugo Basile, Varese, Italy, Model 7750). Rats were given 2 training sessions 10 minutes apart before surgery. Rats were first habituated to the stationary rod. After habituation they were exposed to the rotating rod. The rod was started at 2 rpm and accelerated linearly to 20 rpm within 300 sec. Latency to fall off the rotarod was then determined before ischemia (presurgery) and 24, 48 and 72 h after stroke. Animals were required to stay on the accelerating rod for a minimum of 30 sec. If they were unable to reach this criterion, the trial was repeated for a maximum of five times. The two best (largest) fall latency values a rat could achieve then were averaged and used for data analysis. Rats not falling off within 5 min were given a maximum score of 300 seconds [7]. A sham-operated group was also included (n=10). The investigator performing the neurological evaluation and rotarod test did not know the identity of the experimental groups until completion of data analysis.

### 2.5. Quantification of brain infarct volume

The method for quantification of infarct volume was performed exactly as previously reported [25,66,78]. Briefly, after completing the neurological evaluation and the rotarod test at 3 days after transient FCI, the animals were sacrificed under deep anesthesia and brains were removed, frozen and coronally sectioned into six 2-mm-thick slices (from rostral to caudal, first to sixth). The brain slices were incubated for 30 min in a 2% solution of 2,3,5-triphenyltetrazolium chloride (TTC) (Sigma Chemical Co., Saint Louis, MO, USA) at 37 °C and fixed by immersion in a 4% paraformaldehyde solution in phosphate-buffered saline pH 7.4. Six TTC-stained brain sections per animal were placed directly on the scanning screen of a color flatbed scanner (Hewlett Packard, HP Scanjet 5370 C) within 7 days. Following image acquisition, the images were analyzed blindly using a commercial image processing software program (Photoshop, version 7.0, Adobe Systems; Mountain View, CA, USA). Measurements were made by manually outlining the margins of infarcted areas. The unstained area of the fixed brain section was defined as infarcted. Cortical and subcortical uncorrected infarcted areas and total hemispheric areas were calculated separately for each coronal slice. Total cortical and striatal uncorrected infarct volumes were calculated by multiplying the infarcted area by the slice thickness and summing the volume of the six slices. A corrected infarct volume was calculated to compensate for the effect of brain edema. An edema index was calculated by dividing the total volume of the hemisphere ipsilateral to MCAO by the total volume of the contralateral hemisphere. The actual infarct volume adjusted for edema was calculated by dividing the infarct volume by the edema index [48,59,79].

In this study we measured the size of infarction by TTC staining at 3 days after ischemia. In our pilot studies, we compared the results of infarcted area ($mm^2$) obtained by TTC staining with those obtained using conventional hematoxylin-eosin (H&E) staining. Rats underwent transient FCI (1 h of MCAO) and following 3 days of reperfusion, animals were sacrificed, brains were cut into 2 mm slices and stained with TTC. The cross-sectional area of the ipsilateral (ischemic) hemisphere and the contralateral hemisphere for each brain slice was calculated. H&E-stained sections were also prepared from the brain slices (the same slices previously stained with TTC) and analyzed for cross-sectional area as described above. Measurements obtained from corresponding TTC-stained brain slices and H&E-stained brain sections were directly compared using linear regression and Pearson product moment correlation analysis. TTC and H&E defined area measurements were highly correlated (correlation coefficient = 0.971, P<0.001) (Candelario-Jalil et al., pilot studies), indicating that in our experimental conditions measurement of infarct volume using either TTC or H&E staining gives very similar results even after 3 days of reperfusion. Our results are in line with those previously reported [21,39,54], which show that TTC staining could be used to measure infarct volume even 3-4 days after stroke. TTC staining has several advantages over other histological techniques (as compared to H&E or Nissl staining techniques, TTC staining is cheap, easily to perform and extremely quick).

### 2.6. Effect of nimesulide on PGE$_2$ elevation in the ischemic brain

In separate rats, the effect of nimesulide on the elevation in PGE$_2$ concentrations was studied. PGE$_2$ levels were determined 24 h after induction of stroke, because at this time the concentrations of this prostaglandin were dramatically elevated in the ischemic brain (see Fig. 5, Results section). In addition, a previous study demonstrated a marked elevation in PGE$_2$ at this time point following transient FCI [51]. Nimesulide (12 mg/kg; i.p.; n=9) or vehicle (n=10) were administered at 6, 14 and 22 h after induction of ischemia. Rats were sacrificed for PGE$_2$ analysis 2 h after the last nimesulide treatment following a similar experimental strategy to that employed by Nogawa et al. [51]. A 4-mm-thick coronal brain slice was cut at the level of the optic chiasm, and the infarcted cortex was quickly dissected out on a chilled plate. The corresponding region of the contralateral cortex was also sampled. Tissue was rapidly collected, weighed and frozen in liquid nitrogen. A sham-operated group was included in this experiment (n=7).

### 2.7. Prostaglandin $E_2$ (PGE$_2$) Enzyme Immunoassay

Tissue concentration of $PGE_2$, one of the major cyclooxygenase reaction products in the brain [51], was determined using a commercial enzyme immunoassay kit (RPN 222, Amersham Pharmacia Biotech Inc., Piscataway, NJ, USA) according to the instructions of the manufacturer. The tissue was homogenized in 50 mM Tris-HCl (pH 7.4) and extracted with 100 % methanol [57]. After centrifugation, the supernatant was diluted with acidified 0.1 M phosphate buffer (pH 4.0; final methanol concentration, 15%) and applied to activated octadecylsilyl (ODS)-silica reverse-phase columns (Sep-Pak C18, Waters Associates, Milford, MA, USA). The columns were rinsed with 5 mL of distilled water followed by 5 mL of n-hexane, and $PGE_2$ was eluted twice with 2 mL of ethyl acetate containing 1% methanol. The ethyl acetate fraction was evaporated and resuspended in 1 mL of the buffer provided with the kit. The assay is based on competition between unlabeled $PGE_2$ and a fixed amount of peroxidase-labeled $PGE_2$ for a limited number of binding sites on a $PGE_2$-specific antibody. Briefly, samples or $PGE_2$ standard were incubated, shaken at room temperature for 60 min with specific anti-$PGE_2$ antibody and peroxidase-conjugated $PGE_2$ antibody in a goat anti-mouse IgG-precoated 96-well plate. After washing, 3,3´,5,5´-tetramethylbenzidine substrate was added to the wells and after 30 min the reaction was stopped by adding 1 mM sulphuric acid. Subsequently, absorbance was measured at 450 nm. The detection limit of this assay is 16 pg/mL.

### 2.8. Statistical analysis

Data are presented as means ± S.D. Infarct volumes, infarct areas and $PGE_2$ data were analyzed using t-test (2 groups) or one way ANOVA with *post-hoc* Student-Newman-Keuls test (multiple comparison). Neurological deficit scores were analyzed by Kruskal-Wallis nonparametric ANOVA followed by the Dunn test (multiple comparison) or Mann-Whitney test for analysis of individual differences. Rotarod performance was expressed as a percentage of pre-surgery values for each rat and analyzed by ANOVA for repeated measures followed by the Student-Newman-Keuls test. Differences were considered significant when $p < 0.05$.

## 3. RESULTS

### 3.1. Effects of different doses of nimesulide on stroke-induced neuronal damage

Transient occlusion (1 h) of the middle cerebral artery (MCA) by the intraluminal suture method consistently produced large infarcts in the territory of the MCA, involving both cortical and subcortical structures. Three different clinically-relevant doses of nimesulide were evaluated in this model of transient FCI. Table 2 shows the neuroprotective effects of this compound when administered both as a single dose and in a long-term treatment paradigm. Nimesulide dose-dependently reduced total and cortical infarct volumes and a modest protective effect was also observed with the highest dose (12 mg/kg) in the subcortical areas. As compared with a single dose of nimesulide (16 % of infarct reduction), the highest dose of nimesulide given as repeated treatments for 2 days reduced by 59 % total (cortical and subcortical) infarct volume assessed 3 days after the ischemic stroke. A representative TTC-stained brain section of vehicle and nimesulide-treated animals (12 mg/kg, repeated doses) is shown in Fig. 1.

When considered separately, mean cortical infarct volume was decreased by 64 % by treatment with nimesulide (12 mg/kg; 6 injections for 2 days post-stroke) compared to vehicle rats (81.5 ± 21.2 and 227.6 ± 27.4 mm$^3$, respectively; $p < 0.001$). Mean subcortical infarct volume was only slightly reduced by nimesulide at a dose of 12 mg/kg as shown in Table 2. Thus, the significant reduction in total infarction is chiefly the result of the potent neuroprotective effect of nimesulide in cortical areas following ischemia.

The rostrocaudal distribution of cortical and subcortical infarct areas in the vehicle and nimesulide 12 mg/kg groups is depicted in Fig. 2. Cortical infarct areas were significantly smaller (p<0.05) in nimesulide-treated rats than in the vehicle group at all coronal levels (Fig. 2A). On the contrary of what was found in the case of infarcted cerebral cortex, nimesulide only reduced modestly subcortical infarct areas at coronal levels 2 and 3 (Fig. 2B).

On the other hand, vehicle-treated control rats exhibited significantly higher neurological deficits scores than sham-operated controls (Fig. 3). As compared to vehicle, treatment with nimesulide at the three doses examined significantly (p<0.05) improved neurological outcome at 24, 48 and 72 h after stroke, although rats in the group treated with nimesulide 12 mg/kg had statistically significant better neurological deficit scores than animals treated with the lowest doses at 48 and 72 h after FCI. No significant reduction in neurological deficits was observed when nimesulide (12 mg/kg) was given as a single dose although a trend towards a better neurological outcome was noticed at 24 h post-stroke (data not shown).

In the accelerating rotarod test, each animal acted as its own control, and performance was compared with pre-surgery results (Table 3). Vehicle-treated ischemic rats showed significant impairments in performance 24, 48 and 72 h after stroke compared with sham-operated rats. Repeated administrations of the three doses of nimesulide similarly protected (p<0.05) against motor impairment seen after stroke (Table 3). Treatment with a single dose given immediately after ischemia did not improve performance in the rotarod test (Table 3).

### 3.2. Time window for nimesulide protection in rats subjected to transient focal cerebral ischemia

Given the potent neuroprotection observed with nimesulide at a dose of 12 mg/kg, we decided to select this dose for subsequent experiments evaluating the therapeutic window of protection of this COX-2 inhibitor. When nimesulide treatment was delayed for 3 to 24 h after ischemia, a significant reduction in infarct volume was observed 3 days after stroke, although an overall decline of efficacy with post-treatment time was observed (Table 4). Total infarct volume was reduced by 51%, 42%, 37% and 17% when drug administration began 3, 6, 12 or 24 h after stroke, respectively. Similar to what was found in the case of immediate administration (Table 2), nimesulide did not confer protection against ischemic damage in subcortical areas (Table 4). Thus, the significant reduction in cortical infarction accounts for the marked decrease in total infarct volume observed in rats given nimesulide in a delayed therapeutic schedule.

Interestingly, nimesulide not only reduced infarction but enhanced functional recovery 3 days after ischemia even when its first administration began 24 h after stroke (Fig. 4). Neurological deficits were significantly reduced by post-ischemic treatment with nimesulide (Fig. 4A). Furthermore, latency to remain on the accelerating rotarod was compared in ischemic rats treated with either vehicle or with nimesulide in the delayed administration paradigm. Nimesulide-treated rats showed significantly increased fall latencies compared to those from animals given vehicle alone when treatment was delayed for 3-12 h after ischemia (Fig. 4B). However, this protective effect was lost when the first administration is delayed until 24 h following the ischemic episode (P=0.183, Student's t-test, Fig. 4B).

### 3.3. Effects of nimesulide on $PGE_2$ accumulation in the ischemic brain

To determine whether delayed treatment with nimesulide was effective in reducing $PGE_2$ accumulation in the ischemic brain, $PGE_2$ levels were measured 24 h after transient stroke in animals treated with nimesulide (12 mg/kg; 6 h delayed treatment) or vehicle. There was a dramatic increase (by 244 %) in $PGE_2$ levels in the ischemic cortex in those rats treated with the vehicle as compared to the levels of this prostaglandin in the contralateral cortex (P<0.001; Fig. 5). Nimesulide completely abolished the post-ischemic increases in $PGE_2$ (P<0.001 from vehicle) in the ipsilateral cerebral cortex (stroke side) and the $PGE_2$ concentration in the injured cerebral cortex was not different (P>0.05) from that in the intact side (contralateral) as shown in Fig. 5.

## 4. DISCUSSION

Gone are the days when drugs that confer neuroprotection when given before or a short period after cerebral ischemia can be considered relevant for therapy of ischemic stroke. Several agents from this group have been evaluated clinically and failed. COX-2 inhibition has emerged as a potential therapeutic strategy for cerebral ischemia, targeting critical late-occurring pathophysiological events which exacerbate the initial brain damage triggered by the ischemic episode.

This study demonstrated that the COX-2 inhibitor nimesulide appreciably reduces cerebral infarction, $PGE_2$ accumulation, and also improves functional outcome after transient MCAO in rats. Interestingly, the effects of nimesulide on both histological and functional recovery were evident even when the first administration was delayed up to 24 h after stroke. Although a previous report demonstrated positive effects with a COX-2 selective inhibitor (NS-398) when given after FCI [51], results from the present study demonstrate for the first time the wide therapeutic window for nimesulide in a rat stroke model and more importantly demonstrated that nimesulide also markedly improved functional recovery. As previously suggested [40], pre-clinical studies directed toward demonstrating functional improvement of neurological function in addition to reduction of infarct size may improve the predictive value of animal models for clinical efficacy with novel neuroprotective agents.

Present results that nimesulide protects neurons when administered several hours after stroke is consistent with our previous studies which have found that COX-2 selective inhibitors have a wide therapeutic window for protection in global cerebral ischemia [9-12], thus extending our observations to a model of transient focal ischemic stroke.

It is important to discuss the finding that nimesulide did not reduce damage to subcortical areas (mainly striatum) when administered in a delayed fashion and only slightly diminished infarct volume in striatum when given immediately after stroke at the highest dose (Table 2; Fig. 2). The striatum is considered to be the core of the ischemic lesion, it lacks collateral circulation and has proved relatively refractory to neuroprotection [24]. In addition, results from a previous study [51] indicated that COX-2 protein expression is not upregulated in striatum after transient cerebral ischemia, suggesting that COX-2 seems not to be an important pathophysiological mediator of ischemic damage in this brain region. This probably helps to explain our present results. The mild positive effects seen with the highest dose of nimesulide (Table 2; Fig. 2) might be attributable to other pharmacological effects of this compound not related to COX-2 inhibition, although further studies are needed to support this notion.

On the other hand, repeated treatments with nimesulide afforded a more remarkable neuroprotection than the administration of a single dose given immediately after ischemia (Tables 2 and 3). These findings show the importance of a continuous long-term therapeutic regime after focal stroke in clinical trials to achieve the maximal beneficial effects of neuroprotection with nimesulide to target the delayed progression of tissue damage.

According to our results, the lowest dose of nimesulide (3 mg/kg) reduced neurological deficits and motor impairment (Table 3 and Fig. 3) similarly to the highest dose of this COX-2 inhibitor (12 mg/kg), but these positive effects were not accompanied by a significant reduction in infarct volume (Table 2). This might reflect the fact that unlike ischemic injury to many other tissues, the severity of disability is not predicted well by the amount of brain tissue lost. For example, damage to a small area in the medial temporal lobe may lead to severe disability, while damage to a greater volume elsewhere had little effect on function [22]. The majority of studies directed toward determining neuroprotective efficacy have used reduction of infarct volume as a measure of a drug's efficacy in animals subjected to focal ischemia. Although it is presumed that reduced lesion size will translate to improved functional outcome, a direct correlation is not always observed in animals models [32] or in stroke patients [68].

Pharmacological inhibition of COX-2 has been previously shown to reduce N-methyl-D-aspartate-mediated neuronal cell death both in vitro [29] and in vivo [36]. In addition, recent investigations have found a potentiation of excitotoxicity in transgenic mice overexpressing neuronal COX-2 [37] and a significant reduction in ischemic brain injury in COX-2-deficient mice [36,62].

COX-2 expression by itself does not lead to neuronal death since a variety of healthy neuronal populations throughout the CNS express COX-2 mRNA and protein under normal conditions [5,77] and COX-2 expression can be experimentally induced without causing neuronal death [56]. In the study of Planas et al. [56], they found the same level of COX-2 induction in response to both 10 min (mild enough not to cause inflammation or cell death) and 1 h of stroke (which leads to brain infarct), suggesting that COX-2 would only mediate neuronal injury in the context of an inflammatory response. Results from previous studies indicated a close relationship between COX-2 and iNOS in experimental models of FCI [49,52]. These two pro-inflammatory enzymes are expressed at the same time and in close proximity in penumbral regions [52]. Pharmacological inhibition of iNOS attenuates accumulation of $PGE_2$ in the ischemic brain and COX-2 selective inhibitors decrease ischemic injury in wild type mice but not in iNOS knockout mice [49]. These results suggest that iNOS and COX-2 may work synergistically to exacerbate damage in brain, perhaps through the formation of peroxynitrite and the ensuing oxidative stress.

The production of pro-inflammatory prostanoids is an injurious mechanism associated to the COX-2 enzymatic activity and this process is associated with the generation of highly reactive oxygen species, which have potent deleterious effects on cells [42]. We found that delayed treatment of rats with nimesulide completely abolished the marked increase in $PGE_2$ seen in the ischemic cortex 24 h after stroke. These results are consistent with the hypothesis that COX-2 reaction products contribute to the delayed progression of brain injury following transient FCI.

Several additional mechanisms could account for the neuroprotection conferred by nimesulide in focal stroke. The possibility that nimesulide diminished cerebral injury through mechanisms involving reduction of oxidative damage, inhibition of pro-inflammatory cytokines production and blockade of apoptotic pathways cannot be excluded. Recently, we have found that nimesulide is able to reduce oxidative damage following excitotoxic or ischemic brain injury [8,12]. Furthermore, nimesulide has been proven to inhibit TNF-$\alpha$ production [3] and glutamate-mediated apoptotic damage [45]. Further studies are needed to elucidate other potential mechanisms apart from COX-2 inhibition which contribute to the neuroprotective effects of nimesulide in ischemic stroke.

The inflammatory cascade following cerebral ischemia comprises several mediators (e.g., cytokines, chemokines, adhesion molecules, eicosanoids, nitric oxide), which interact among them to produce a long-lasting inflammatory reaction observed not only in animal models but in patients with ischemic stroke [19,35,60]. Inflammation is an attractive pharmacologic opportunity, considering its rapid initiation and progression over many hours/days after stroke and its well-demonstrated contribution to evolution of tissue damage [4]. Nevertheless, anti-inflammatory interventions have been shown to interfere with nervous regeneration/plasticity and recovery following some types of neuronal injury [17,31], suggesting that the potential beneficial effects of inflammation in tissue repair and remodeling need to be considered when developing treatment strategies aimed at reducing post-ischemic inflammation.

Considering all these previous evidences on the dual role of inflammation following brain injury, further work needs to be done in order to investigate the long-term effects of nimesulide in the ischemic brain before this compound could be used in the treatment of patients suffering from ischemic stroke.

In summary, the present study has demonstrated a marked neuroprotective effect of nimesulide against transient focal ischemic injury at therapeutically relevant doses when administered even 24 h after ischemia. Of great importance is the result that nimesulide not only reduced infarct size but also improved functional outcome. Our findings hold a therapeutic promise to intervene neuronal injury evolving after stroke with the COX-2 inhibitor nimesulide.

## REFERENCES

[1]    K. Abe, Therapeutic potential of neurotrophic factors and neural stem cells against ischemic brain injury, J. Cereb. Blood Flow Metab. 20 (2000) 1393-1408.

[2]    G. W. Albers, Expanding the Window for Thrombolytic Therapy in Acute Stroke, Stroke 30 (1999) 2230-2237.

[3]    A. Azab, V. Fraifeld, J. Kaplanski, Nimesulide prevents lipopolysaccharide-induced elevation in plasma tumor necrosis factor-alpha in rats, Life Sci. 63 (1998) 323-327.

[4]    F.C. Barone, G.Z. Feuerstein, Inflammatory mediators and stroke: new opportunities for novel therapeutics, J. Cereb. Blood Flow Metab. 19 (1999) 819-834.

[5]    C.D. Breder, D. Dewitt, R.P. Kraig, Characterization of inducible cyclooxygenase in rat brain, J. Comp. Neurol. 355 (1995) 296-315.

[6]    J.K. Callaway, M.J. Knight, D.J. Watkins, P.M. Beart, B. Jarrott, Delayed treatment with AM-36, a novel neuroprotective agent, reduces neuronal damage after endothelin-1-induced middle cerebral artery occlusion in conscious rats, Stroke 30 (1999) 2704–2712.

[7]    J.K. Callaway, A.J. Lawrence, B. Jarrott, AM-36, a novel neuroprotective agent, profoundly reduces reactive oxygen species formation and dopamine release in the striatum of conscious rats after endothelin-1-induced middle cerebral artery occlusion, Neuropharmacology 44 (2003) 787–800.

[8]    E. Candelario-Jalil, H.H. Ajamieh, S. Sam, G. Martínez, O.S. León, Nimesulide limits kainate-induced oxidative damage in the rat hippocampus, Eur. J. Pharmacol. 390 (2000) 295-298.

[9]    E. Candelario-Jalil, D. Alvarez, A. González-Falcón, M. García-Cabrera, G. Martínez-Sánchez, N. Merino, A. Giuliani, O.S. León, Neuroprotective efficacy of nimesulide against hippocampal neuronal damage following transient forebrain ischemia, Eur. J. Pharmacol. 453 (2002) 189-195.

[10]    E. Candelario-Jalil, D. Alvarez, J.M. Castañeda, S.M. Al-Dalain, G. Martínez, N. Merino, O.S. León, The highly selective cyclooxygenase-2 inhibitor DFU is neuroprotective when given several hours after transient cerebral ischemia in gerbils, Brain Res. 927 (2002) 212-215.

[11]    E. Candelario-Jalil, A. González-Falcón, M. García-Cabrera, D. Alvarez, S.M. Al-Dalain, G. Martínez, O.S. León, J.E. Springer, Assessment of the relative contribution of COX-1 and COX-2 isoforms to ischemia-induced oxidative damage and neurodegeneration following transient global cerebral ischemia, J. Neurochem. 86 (2003) 545–555.

[12]    E. Candelario-Jalil, D. Alvarez, N. Merino, O.S. León, Delayed treatment with nimesulide reduces measures of oxidative stress following global ischemic brain injury in gerbils, Neurosci. Res. 47 (2003) 245-253.

[13]    P.E. Chabrier, M. Auguet, B. Spinnewyn, S. Auvin, S. Cornet, C. Demerlé-Pallardy, C. Guilmard-Favre, J.G. Marin, B. Pignol, V. Gillard-Roubert, C. Roussillot-Charnet, J. Schulz, I. Viossat, D. Bigg, S. Moncada, BN 80933, a dual inhibitor of neuronal nitric oxide synthase and lipid peroxidation: a promising neuroprotective strategy, Proc. Natl. Acad. Sci. USA 96 (1999) 10824-10829.

[14]    J.A. Clemens, Cerebral ischemia: gene activation, neuronal injury, and the protective role of antioxidants, Free Radic. Biol. Med. 28 (2000) 1526-1531.

[15]    Y. Collaço-Moraes, B. Aspey, M. Harrison, J. de Belleroche, Cyclo-oxygenase-2 messenger RNA induction in focal cerebral ischemia, J. Cereb. Blood Flow Metab. 16 (1996) 1366-1372.

[16]    L. Cullen, L. Kelly, S.O. Connor, D.J. Fitzgerald, Selective cyclooxygenase-2 inhibition by nimesulide in man, J. Pharmacol. Exp. Ther. 287 (1998) 578-582.

[17]    P.K. Dash, S.A. Mach, A.N. Moore, Regional expression and role of cyclooxygenase-2 following experimental traumatic brain injury, J. Neurotrauma 17 (2000) 69-81.

[18]    G.J. del Zoppo, Clinical trials in acute stroke: why have they not been successful?, Neurology 51 (1998) S59-S61.

[19]    G. del Zoppo, I. Ginis, J.M. Hallenbeck, C. Iadecola, X. Wang, G.Z. Feuerstein, Inflammation and stroke: putative role for cytokines, adhesion molecules and iNOS in brain response to ischemia, Brain Pathol. 10 (2000) 95-112.

[20]  W. Derk, M. Krieger, A. Abou-Chebl, J.C. Andrefsky, C.A. Sila, I. L. Katzan, M.R. Mayberg, A.J. Furlan, Cooling for Acute Ischemic Brain Damage (COOL AID): An Open Pilot Study of Induced Hypothermia in Acute Ischemic Stroke, Stroke 32 (2001) 1847-1854.

[21]  S. Dore, T. Otsuka, T. Mito, N. Sugo, T. Hand, L. Wu, P.D. Hurn, R.J. Traystman, K. Andreasson, Neuronal expression of cyclooxygenase-2 increases stroke damage, Ann. Neurol. 54 (2003) 155-162.

[22]  L.L. Dugan, D.W. Choi, Hypoxic-ischemic brain injury and oxidative stress. In: G.J. Siegel, B.W. Agranoff, R.W. Albers, S.K. Fisher, M.D. Uhler (Eds.), Basic Neurochemistry: Molecular, Cellular and Medical Aspects. Lippincott-Raven Publishers, Philadelphia, 1999, pp. 711-729.

[23]  A.G. Dyker, K.R. Lees, Duration of neuroprotective treatment for ischemic stroke, Stroke 29 (1998) 535-542.

[24]  M. Fisher, J.H. Garcia, Evolving stroke and the ischemic penumbra, Neurology 47 (1996) 884-888.

[25]  A. González-Falcón, E. Candelario-Jalil, M. García-Cabrera, O.S. León, Effects of pyruvate administration on infarct volume and neurological deficits following permanent focal cerebral ischemia in rats, Brain Res. 990 (2003) 1-7.

[26]  S.H. Graham, J. Chen, Programmed cell death in cerebral ischemia, J. Cereb. Blood Flow Metab. 21 (2001) 99-109.

[27]  W. Hacke, M. Kaste, C. Fieschi, D. Toni, E. Lesaffre, R. von Kummer, G. Boysen, E. Bluhmki, G. Hoxter, M.H. Mahagne, Intravenous thrombolysis with recombinant tissue plasminogen activator for acute hemispheric stroke: the European Cooperative Acute Stroke Study (ECASS), JAMA 274 (1995) 1017-1025.

[28]  W. Hacke, T. Brott, L. Caplan, D. Meier, C. Fieschi, R. von Kummer, G. Donnan, W.D. Heiss, N.G. Wahlgren, M. Spranger, G. Boysen, J.R. Marler, Thrombolysis in acute ischemic stroke: controlled trials and clinical experience, Neurology 53 (1999) S3-S14.

[29]  S.J. Hewett, T.F. Uliasz, A.S. Vidwans, J.A. Hewett, Cyclooxygenase-2 contributes to N-methyl-D-aspartate-mediated neuronal cell death in primary cortical cell culture, J. Pharmacol. Exp. Ther. 293 (2000) 417-425.

[30]  S.L. Hickenbottom, J. Grotta, Neuroprotective therapy, Semin. Neurol. 18 (1998) 485-492.

[31]  D.L. Hirschberg, E. Yoles, M. Belkin, M. Schwartz, Inflammation after axonal injury has conflicting consequences for recovery of function: rescue of spared axons is impaired but regeneration is supported, J. Neuroimmunol. 50 (1994) 9-16.

[32]  A.J. Hunter, K.B. Mackay, D.C. Rogers, To what extent have functional studies of ischaemia in animals been useful in the assessment of potential neuroprotective agents?, Trends Pharmacol. Sci. 19 (1998) 59-66.

[33]  A.J. Hunter, J. Hatcher, D. Virley, P. Nelson, E. Irving, S.J. Hadingham, A.A. Parsons, Functional assessments in mice and rats after focal stroke, Neuropharmacology 39 (2000) 806–816.

[34]  C. Iadecola, M. Alexander, Cerebral ischemia and inflammation, Curr. Opin. Neurol. 14 (2001) 89-94.

[35]  C. Iadecola, C. Forster, S. Nogawa, H.B. Clark, M.E. Ross, Cyclooxygenase-2 immunoreactivity in the human brain following cerebral ischemia, Acta Neuropathol. 98 (1999) 9-14.

[36]  C. Iadecola, K. Niwa, S. Nogawa, X. Zhao, M. Nagayama, E. Araki, S. Morham, M. Ross, Reduced susceptibility to ischemic brain injury and NMDA-mediated neurotoxicity in cyclooxygenase-2-deficient mice, Proc. Natl. Acad. Sci. USA 98 (2001) 1294-1299.

[37]  K.A. Kelley, L. Ho, D. inger, J. Freire-Moar, C.B. Borelli, P.S. Aisen, G.M. Pasinetti, Potentiation of excitotoxicity in transgenic mice overexpressing neuronal cyclooxygenase-2, Am. J. Pathol. 155 (1999) 995-1004.

[38]  J. Koistinaho, S. Koponen, P.H. Chan, Expression of cyclooxygenase-2 mRNA after global ischemia is regulated by AMPA receptors and glucocorticoids, Stroke 30 (1999) 1900-1906.

[39]  J.M. Lee, G.J. Zipfel, K.H. Park, Y.Y. He, C.Y. Hsu, D.W. Choi, Zinc translocation accelerates infarction after mild transient focal ischemia, Neuroscience 115 (2002) 871-878.

[40]  Q. Li, D. Stephenson, Postischemic administration of basic fibroblast growth factor improves sensorimotor function and reduces infarct size following permanent focal cerebral ischemia in the rat, Exp. Neurol. 177 (2002) 531-537.

[41]  E.Z. Longa, P.R. Weinstein, S. Carlson, R. Cummins, Reversible middle cerebral artery occlusion without craniectomy in rats, Stroke 20 (1989) 84-91.

[42]  L. Marnett, S. Rowlinson, D. Goodwin, A. Kalgutkar, C. Lanzo, Arachidonic acid oxygenation by COX-1 and COX-2. Mechanisms of catalysis and inhibition, J. Biol. Chem. 274 (1999) 22903-22906.

[43]  S. Miettinen, F.R. Fusco, J. Yrjanheikki, R. Keinanen, T. Hirvonen, R. Roivainen, M. Narhi, T. Hokfelt, J. Koistinaho, Spreading depression and focal brain ischemia induce cyclooxygenase-2 in cortical neurons through N-methyl-D-aspartic acid-receptors and phospholipase A2, Proc. Natl. Acad. Sci. USA 94 (1997) 6500-6505.

[44]  K. Minematsu, L. Li, C.H. Sotak, M.A. Davis, T. Fisher, Reversible focal ischemic injury demonstrated by diffusion-weighted magnetic resonance imaging in rats, Stroke 23 (1992) 1304-1310.

[45]  M. Mirjany, L. Ho, G.M. Pasinetti, Role of cyclooxygenase-2 in neuronal cell cycle activity and glutamate-mediated excitotoxicity, J. Pharmacol. Exp. Ther. 301 (2002) 494-500.

[46]  M. Modo, P. Rezaie, P. Heuschling, S. Patel, D.K. Male, H. Hodges, Transplantation of neural stem cells in a rat model of stroke: assessment of short-term graft survival and acute host immunological response, Brain Res. 958 (2002) 70-82.

[47]  N. J. Mulcahy, J. Ross, N.J. Rothwell, S.A. Loddick, Delayed administration of interleukin-1 receptor antagonist protects against transient cerebral ischaemia in the rat, Br. J. Pharmacol. 140 (2003) 471–476.

[48]  N. Nagai, M. de Mol, B. van Hoef, M. Verstreken, M. Collen, Depletion of circulating alpha(2)-antiplasmin by intravenous plasmin or immunoneutralization reduces focal cerebral ischemic injury in the absence of arterial recanalization, Blood 97 (2001) 3086-3092.

[49]  M. Nagayama, K. Niwa, T. Nagayama, M.E. Ross, C. Iadecola, The cyclooxygenase-2 inhibitor NS-398 ameliorates ischemic brain injury in wild-type mice but not in mice with deletion of the inducible nitric oxide synthase gene, J. Cereb. Blood Flow Metab. 19 (1999) 1213-1219.

[50]   M. Nakayama, K. Uchimura, R.L. Zhu, T. Nagayama, M.E. Rose, R.A. Stetler, P.C. Isakson, J. Chen, S.H. Graham, Cyclooxygenase-2 inhibition prevents delayed death of CA1 hippocampal neurons following global ischemia, Proc. Natl. Acad. Sci. USA 95 (1998) 10954-10959.

[51]   S. Nogawa, F. Zhang, M.E. Ross, C. Iadecola, Cyclo-oxygenase-2 gene expression in neurons contributes to ischemic brain damage, J. Neurosci. 17 (1997) 2746-2755.

[52]   S. Nogawa, C. Forster, F. Zhang, M. Nagayama, M.E. Ross, C. Iadecola, Interaction between inducible nitric oxide synthase and cyclooxygenase-2 after cerebral ischemia, Proc. Natl. Acad. Sci. USA 95 (1998) 10966-10971.

[53]   B. Onténiente, C. Couriaud, J. Braudeau, A. Benchoua, C. Guégan, The mechanisms of cell death in focal cerebral ischemia highlight neuroprotective perspectives by anti-caspase therapy, Biochem. Pharmacol. 66 (2003) 1643–1649.

[54]   S. Parmentier, G.A. Bohme, D. Lerouet, D. Damour, J.M. Stutzmann, I. Margaill, M. Plotkine, Selective inhibition of inducible nitric oxide synthase prevents ischaemic brain injury, Br. J. Pharmacol. 127 (1999) 546-552.

[55]   A.M. Planas, M.A. Soriano, E. Rodríguez-Farré, I. Ferrer, Induction of cyclooxygenase-2 mRNA and protein following transient focal ischemia in the rat brain, Neurosci. Lett. 200 (1995) 187-190.

[56]   A.M. Planas, M.A. Soriano, C. Justicia, E. Rodríguez-Farré, Induction of cyclooxygenase-2 in the rat brain after a mild episode of focal ischemia without tissue inflammation or neural cell damage, Neurosci Lett. 275 (1999) 141-144.

[57]   W.S. Powell, Rapid extraction of arachidonic acid metabolites from biological samples using octadecylsilyl silica. In: Lands, W.E.M., Smith, W.L., (Eds), Methods in Enzymology. Academic Press, Orlando, 1982, pp. 466-477.

[58]   V.L.R. Rao, A. Dogan, K.G. Todd, K.K. Bowen, B.T. Kim, J.D. Rothstein, R.J. Dempsey, Antisense knockdown of the glial glutamate transporter GLT-1, but not the neuronal glutamate transporter EAAC1, exacerbates transient focal cerebral ischemia-induced neuronal damage in rat brain, J. Neurosci. 21 (2001) 1876-1883.

[59]   D. Reglodi, A. Somogyvari-Vigh, S. Vigh, T. Kozicz, A. Arimura, Delayed systemic administration of PACAP38 is neuroprotective in transient middle cerebral artery occlusion in the rat, Stroke 31 (2000) 1411-1417.

[60]   T. Sairanen, A. Ristimaki, A. Paetau, P.J. Lindsberg, Cyclooxygenase-2 is induced globally in infarcted human brain, Ann. Neurol. 43 (1998) 738-747.

[61]   O. Sanz, A. Estrada, I. Ferrer, A.M. Planas, Differential cellular distribution and dynamics of HSP70, cyclooxygenase-2, and c-Fos in the rat brain after transient focal ischemia or kainic acid, Neuroscience 80 (1997) 221-232.

[62]   T. Sasaki, K. Kitagawa, K. Yamagata, T. Takemiya, S. Tanaka, E. Omura-Matsuoka, S. Sugiura, M. Matsumoto, M. Hori, Amelioration of hippocampal neuronal damage after transient forebrain ischemia in cyclooxygenase-2-deficient mice, J. Cereb. Blood Flow Metab. 24 (2004) 107–113.

[63]   S.I. Savitz, D.M. Rosenbaum, J.H. Dinsmore, L.R. Wechsler, L.R. Caplan LR, Cell transplantation for stroke, Ann. Neurol. 52 (2002) 266-275.

[64]   R. Schmid-Elsaesser, E. Hungerhuber, S. Zausinger, A. Baethmann, H. J. Reulen, Neuroprotective efficacy of combination therapy with two different antioxidants in rats subjected to transient focal ischemia, Brain Res. 816 (1999) 471-479.

[65]   S. Schwab, S. Schwarz, M. Spranger, E. Keller, M. Bertram, W. Hacke, Moderate hypothermia in the treatment of patients with severe middle cerebral artery infarction, Stroke 29 (1998) 2461-2466.

[66]   A. Shuaib, C.X. Wang, T. Yang, R. Noor, Effects of nonpeptide V(1) vasopressin receptor antagonist SR-49059 on infarction volume and recovery of function in a focal embolic stroke model, Stroke 33 (2002) 3033-3037.

[67]   A.L. Sirén, M. Fratelli, M. Brines, C. Goemans, S. Casagrande, P. Lewczuk, S. Keenan, C. Gleiter, C. Pasquali, A. Capobianco, T. Mennini, R. Heulmann, A. Cerami, H. Ehrenreich, P. Ghezzi, Erythropoietin prevents neuronal apoptosis after cerebral ischemia and metabolic stress, Proc. Natl. Acad. Sci. USA 98 (2001) 4044-4049.

[68]   Stroke Therapy Academy Industry Roundtable (STAIR II), Recommendation for clinical trial evaluation of acute stroke therapies, Stroke 32 (2001) 1598-1606.

[69]   Y. Taniguchi, K. Yokoyama, K. Noda, Inhibition of brain cyclooxygenase-2 activity and the antipyretic action of nimesulide, Eur. J. Pharmacol. 330 (1997) 221-229.

[70]   The National Institute of Neurological Disorders and Stroke rt-PA Stroke Study Group, Tissue plasminogen activator for acute ischemic stroke, New England J. Med. 333 (1995) 1581-1587.

[71]   P.L. Toutain, C. Cester, T. Haak, S. Metge, Pharmacokinetic profile and in vitro selective cyclooxygenase-2 inhibition by nimesulide in the dog, J. Vet. Pharmacol. Ther. 24 (2001) 35-42.

[72]   M. van Lookeren Campagne, H. Thibodeaux, N. Bruggen, B. Cairns, R. Gerlai, J.T. Palmer, S.P. Williams, D.G. Lowe, Evidence for a protective role of metallothionein-1 in focal cerebral ischemia, Proc. Natl. Acad. Sci. USA, 96 (1999) 12870–12875.

[73]   J. Vaughan, N. Delanty, Neuroprotective properties of statins in cerebral ischemia and stroke, Stroke 30 (1999) 1969-1973.

[74]   J. Wang, X. Yang, C.V. Camporesi, Z. Yang, G. Bosco, C. Chen, E.M. Camporesi, Propofol reduces infarct size and striatal dopamine accumulation following transient middle cerebral artery occlusion: a microdialysis study, Eur. J. Pharmacol. 452 (2002) 303-308.

[75]   J.H. Weiss, S.L. Sensi, J.Y. Koh, $Zn^{2+}$: a novel ionic mediator of neural injury in brain disease, Trends Pharmacol. Sci. 21 (2000) 395-401.

[76]   A.J. Williams, F.C. Tortella, Neuroprotective effects of the sodium channel blocker RS100642 and attenuation of ischemia-induced brain seizures in the rat, Brain Res. 932 (2002) 45-55.

[77]   K. Yamagata, K.I. Andreasson, P.F. Worley, Expression of a mitogen-inducible cyclooxygenase in brain neurons, regulation by synaptic activity and glucocorticoids, Neuron 11 (1993) 371-386.

[78]   Y. Yang, A. Shuaib, Q. Li, Quantification of infarct size on focal cerebral ischemia model of rats using a simple and economical method, J. Neurosci. Meth. 84 (1998) 9-16.

[79]   Y. Yang, Q. Li, H. Miyashita, W. Howlett, M. Siddiqui, A. Shuaib, Usefulness of postischemic thrombolysis with or without neuroprotection in a focal embolic model of cerebral ischemia, J. Neurosurg. 92 (2000) 841-847.

[80]   M.A. Yenari, S.L. Fink, G.H. Sun, L.K. Chang, M.K. Patel, D.M. Kunis, D. Onley, D.Y. Ho, R.M. Sapolsky, G.K. Steinberg, Gene therapy with HSP72 is neuroprotective in rat models of stroke and epilepsy, Ann. Neurol. 44 (1998) 584-591.

[81]   M.A. Yenari, H. Zhao, R.G. Giffard, R.A. Sobel, R.M. Sapolsky, G.K. Steinberg, Gene therapy and hypothermia for stroke treatment, Ann. NY Acad. Sci. 993 (2003) 54-81.

## TABLE 1

**Treatment schedules to investigate the therapeutic window of nimesulide protection in focal ischemic stroke.**

| Treatment | Post-ischemic time at which rats received the first treatment | Additional treatments (hours after stroke) |
|---|---|---|
| Vehicle (n= 9-10) or Nimesulide 12 mg/kg; i.p (n=9-11) | 3 h | 9, 15, 27, 39 and 51 h |
| | 6 h | 12, 18, 30, 42 and 54 h |
| | 12 h | 18, 24, 36, 48 and 60 h |
| | 24 h | 30, 36, 48 and 60 h |

## TABLE 2

**Dose-response effect of nimesulide in a transient model of focal cerebral ischemia and different effects of a long-term treatment paradigm versus a single dose.**

| Treatment | Total infarct volume ($mm^3$) | Cortical infarct volume ($mm^3$) | Subcortical infarct volume ($mm^3$) |
|---|---|---|---|
| *Repeated doses* | | | |
| Vehicle (n=11) | $289.7 \pm 39.8$ | $227.6 \pm 27.4$ | $58.3 \pm 19.5$ |
| Nimesulide 3 mg/kg (n=9) | $255.2 \pm 40.9$ | $207.3 \pm 22.2$ | $54.5 \pm 12.8$ |
| Nimesulide 6 mg/kg (n=8) | $221.4 \pm 34.5$ [*] | $174.7 \pm 19.3$ [*] | $49.2 \pm 16.4$ |
| Nimesulide 12 mg/kg (n=10) | $119.3 \pm 37.8$ [**, ‡] | $81.5 \pm 21.2$ [**, ‡] | $41.8 \pm 18.6$ [&] |
| *Single dose* | | | |
| Vehicle, single dose (n=8) | $295.2 \pm 42.9$ | $235.9 \pm 30.4$ | $62.3 \pm 20.1$ |
| Nimesulide 12 mg/kg, single dose (n=9) | $247.5 \pm 41.2$ [†] | $194.3 \pm 17.3$ [†] | $51.1 \pm 17.5$ |

Data are mean $\pm$ S.D. [*] P<0.01 compared to vehicle. [**] P<0.001 compared to vehicle. [‡] P<0.05 compared to nimesulide 6 mg/kg. [†] p<0.05 compared to vehicle (single dose). ANOVA followed by Student-Newman-Keuls post-hoc test. [&] P<0.05 compared to vehicle (Student t-test).

## TABLE 3

**Effects of different doses of nimesulide on rotarod performance after middle cerebral artery occlusion in rats.**

| | Rotarod Performance (% of Pre-surgery values) | | |
|---|---|---|---|
| Treatment | 24 h | 48 h | 72 h |
| *Repeated treatments* | | | |
| Sham (n=10) | 121 ± 16 | 139 ± 21 | 128 ± 26 |
| Vehicle (n=11) | 58 ± 17 | 51 ± 20 | 41 ± 19 |
| Nimesulide 3 mg/kg (n=9) | 81 ± 25 * | 87 ± 12 * | 76 ± 21 * |
| Nimesulide 6 mg/kg (n=8) | 86 ± 23 * | 92 ±15 * | 85 ± 16 * |
| Nimesulide 12 mg/kg (n=10) | 87 ± 13 * | 89 ± 14 * | 91 ± 18 * |
| *Single dose* | | | |
| Vehicle, single dose (n=8) | 52 ± 26 | 40 ± 18 | 43 ± 22 |
| Nimesulide 12 mg/kg, single dose (n=9) | 64 ± 15 | 59 ± 21 | 50 ± 28 |

Data are mean ± S.D. for rotarod performance expressed as a percentage of pre-surgery performance for each individual rat. * $P < 0.05$ compared with vehicle at the indicated time. ANOVA followed by Student-Newman-Keuls post-hoc test.

## TABLE 4

**Effects of nimesulide (12 mg/kg; i.p.) or vehicle at different times of administration after transient focal cerebral ischemia on total, cortical and subcortical infarct volumes.**

| | Infarct volume (mm$^3$) | | | | | |
|---|---|---|---|---|---|---|
| | Total | | Cortical | | Subcortical | |
| Time after stroke (h) | Vehicle | Nimesulide | Vehicle | Nimesulide | Vehicle | Nimesulide |
| 3 | 295.7 ± 46.4 (n=9) | 143.4 ± 37.6 ** (n=10) | 237.3 ± 31.2 (n=9) | 93.3 ± 27.5 ** (n=10) | 55.3 ± 18.5 (n=9) | 48.2 ± 21.8 (n=10) |
| 6 | 280.9 ± 31.8 (n=10) | 162.6 ± 42.3 ** (n=10) | 222.8 ± 21.8 (n=10) | 107.7 ± 20.2 ** (n=10) | 60.2 ± 20.4 (n=10) | 57.2 ± 15.3 (n=10) |
| 12 | 299.1 ± 39.4 (n=10) | 188.3 ± 44.8 ** (n=11) | 241.7 ± 29.1 (n=10) | 130.9 ± 24.3 ** (n=11) | 52.9 ± 19.4 (n=10) | 61.2 ± 17.5 (n=11) |
| 24 | 290.5 ± 45.2 (n=9) | 240.2 ± 39.5 * (n=9) | 246.5 ± 39.2 (n=9) | 189.1 ± 28.7 ** (n=9) | 47.2 ± 15.2 (n=9) | 55.1 ± 19.8 (n=9) |

Values are volume of infarction (mean ± S.D.), expressed in cubic millimeters. *$P < 0.05$ and **$P < 0.01$ compared with vehicle control group (Student's t-test).

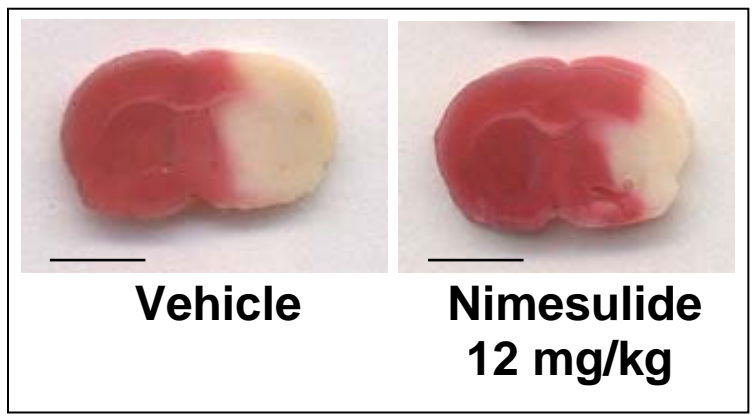

**Vehicle**          **Nimesulide 12 mg/kg**

**Fig. 1**. Representative TTC-stained section of vehicle and nimesulide-treated animals (12 mg/kg; i.p.; repeated doses starting immediately after stroke). Bar= 1 cm.

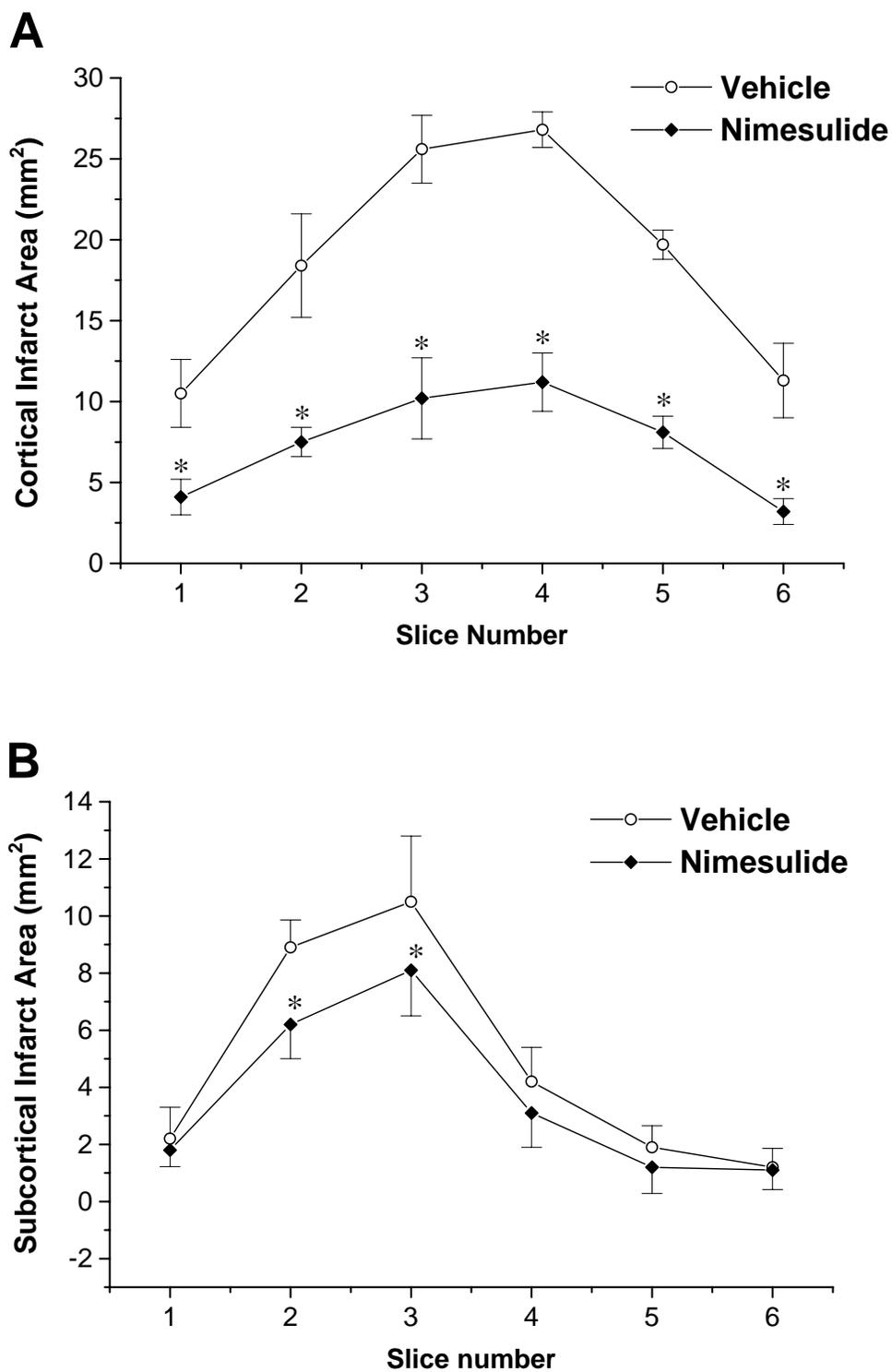

**Fig. 2.** Rostrocaudal distribution of areas of cortical (**A**) and subcortical (**B**) infarction at six coronal levels in nimesulide-treated (12 mg/kg; i.p.; repeated doses starting immediately after stroke) and vehicle-treated rats assessed 3 days after 1 h of transient focal cerebral ischemia. *P<0.05 compared to vehicle.

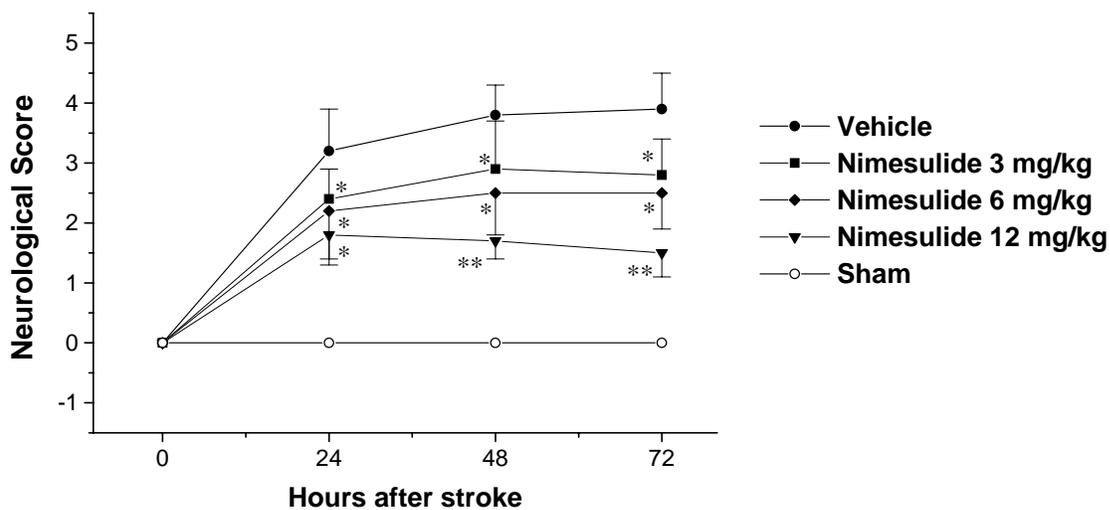

**Fig. 3.** Neurological deficit score 24, 48, and 72 hours after ischemia in sham-operated control group, ischemic vehicle-treated rats and after administration of different doses of nimesulide (repeated treatments for 2 days after stroke). *P<0.05 and **P<0.01 compared with vehicle.

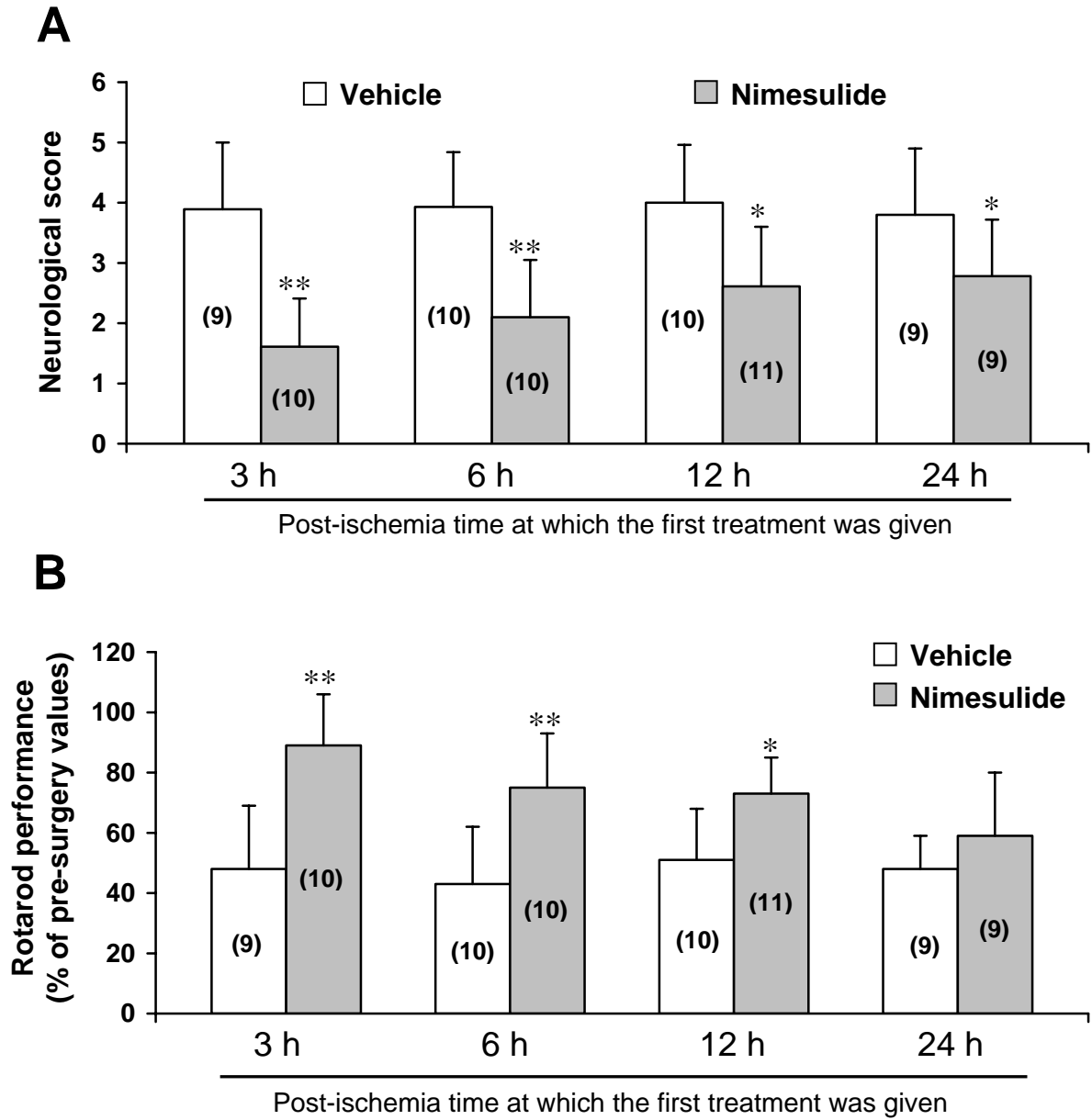

**Fig. 4.** Effect of delayed administration of nimesulide (12 mg/kg; i.p.) on neurological deficit score (**A**) and rotarod performance (**B**) after transient middle cerebral artery occlusion in male Sprague-Dawley rats. Vehicle or nimesulide was administered 3, 6, 12 or 24 h after 1 h of focal ischemia. *P<0.05 and **P<0.01 compared to vehicle. The number in parentheses shows the number of rats used per group.

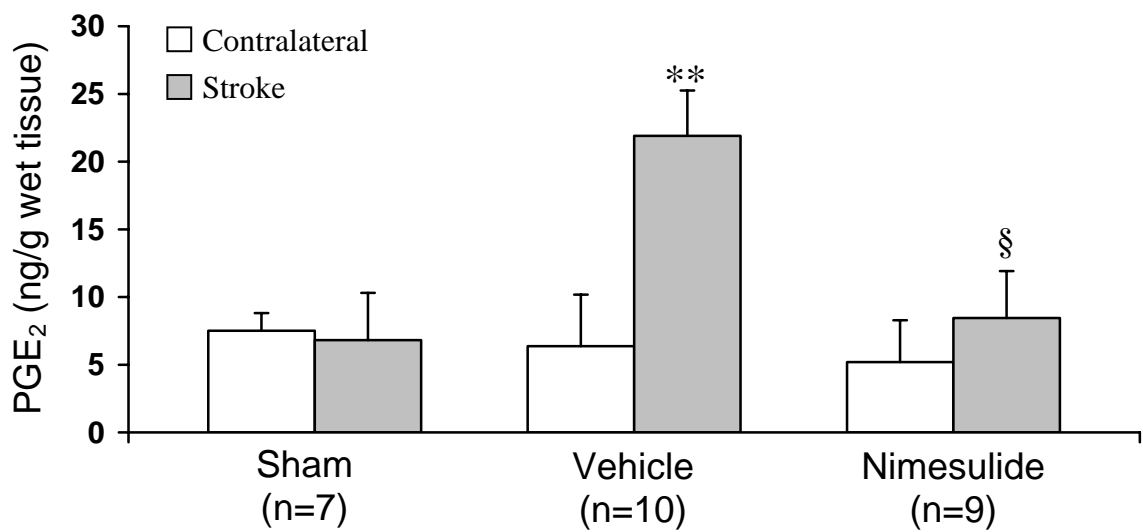

**Fig. 5.** Effect of the COX-2 inhibitor nimesulide (12 mg/kg; i.p.; starting 6 h after ischemia) on postischemic increase in prostaglandin $E_2$ (PGE$_2$) in the ischemic cerebral cortex after 24 h of transient (1 h) focal stroke. Nimesulide completely abolished PGE$_2$ accumulation in the ischemic brain. **P<0.001 from vehicle contralateral. §P<0.001 from vehicle stroke.